\documentstyle[aps,preprint]{revtex}
\begin{document}

\baselineskip=20pt

\thispagestyle{empty}

\vspace{10mm}

\begin{center}
\vspace{5mm}
{\Large \bf GCP Code Applied in $J/\psi$ Production and Absorption}

\vspace{1.5cm}

{\sf   
Xiao-Xia Yao$^{a,b}$, Wei-Qin Chao$^{a,b}$ and Yang Pang$^c$ }\\[2mm]

{\small $^{a}$ CCAST (World Laboratory), Beijing, 100080, P.R. China }\\
{\small $^{b}$ Institute of High Energy Physics, 
Academia Sinica, P.O.Box 918-4, 100039, P.R.China}\\
{\small $^{c}$ Phys Dept. Columbia Univ., New York, USA}\\[10mm]
\end{center}
\begin{abstract}
Using the General Cascade Program(GCP), the production and absorption of 
$J/\psi$ in p-A and A-A collisions have been studied. Nucleon absorption
mechanism and comover absorption mechanism are considered to investigate
the $J/\psi$ suppression. The results agree well with experimental 
data of $J/\psi$ production, except for the data in Pb-Pb collision.
\end{abstract}
\vspace{10mm}
PACS number(s): 24.10.Lx,  25.75.Dw, 12.38.Mh

\vspace{5mm}
\newpage

\centerline{\large{\bf  I. Introduction}}

\medskip

Suppression of $J/\psi$ production in high energy heavy ion collisions 
was proposed as a promising signature for the formation of quark gluon
plasma(QGP) by Matsui and Satz ten years ago\cite{Matsui}.
This suppression effect was observed by NA38 collaboration 
later\cite{NA38}. However, it has been found that $J/\psi$ suppression
exists also in p-A collisions where QGP formation is not possible\cite{mora}. 
Since then, the source of the observed $J/\psi$ suppression has
remained controversial. Recently the NA50 collaboration reported the
anomalous $J/\psi$ suppression in Pb-Pb collision\cite{goni,NA50}. 
Now the debate has 
focussed on whether or not the data can be consistently described 
by hadronic mechanisms\cite{qm97,blaiz,gavin,taian,kharz}. 
So it is important to study first the $J/\psi$ 
suppression caused by hadronic mechanisms.

There are several sources of $J/\psi$ suppression in hadronic matter.
Such as the absorption of $J/\psi$ in nuclear matter, the interaction of
$J/\psi$ with the produced mesons (called comover), gluon 
shadowing in nuclei, energy degradation of produced $J/\psi$.
In this paper, we will discuss the effects of the first two hadronic 
mechanisms on $J/\psi$ suppression.

General Cascade Program(GCP)\cite{high} is designed by Y. Pang. It is a Monte
Carlo simulation code based on cascade method. Cascade method has been 
used extensively in studying relativistic heavy ion collisions. It is
perhaps the only quantitative tool currently available which is capable 
to provide both the overall features and the specific properties of 
nucleus-nucleus collisions at extremely high energies. 

In principle, the cascade algorithm itself is rather simple. One could 
visualize most cascades as collisions of classical billiard balls 
traveling at relativistic speed. Most cascade codes, on the other hand, 
are often quite complex. This is because each code must include a large 
number of physical processes in order to generate a realistic 
nucleus-nucleus collision event. Additional assumptions is often 
required in order
to simulate an actual nucleus-nucleus collision. There are also 
parameters introduced during the cascade. The final result of a
cascade often depends on these assumptions and parameters. 
The physical processes in these models can
be quite different. For most cascade based models, it is often difficult 
to extract the key physics ingredient without having to go over the 
entire source code. It is not easy to compare various models, even 
when good agreements between the model predictions and the 
experimental data in the observed spectra have been reached.  

GCP is designed such that the cascade algorithm is separate from the 
complexities of physics models. So the problem of the cascade can be 
isolated and studied separately, and different physics models can be 
compared in a common framework. GCP itself is only a cascade model,  
a tool for building relativistic cascade models including various 
physical contents. This is the 
key difference between GCP and most other cascade codes.

By now there are many discussion on $J/\psi$ suppression. However, only a 
few works started to treat the Monte Carlo simulation of $J/\psi$ 
production and absorption directly\cite{cass,sai}. While \cite{cass} 
claimed that the $J/\psi$ suppressions in p-A and A-A collisions, 
including Pb-Pb data, could be explained based on 
nucleon and comover absorption, ref\cite{sai} showed that it is impossible 
to fit the abnormal $J/\psi$ suppression in Pb-Pb collision only 
based on hadronic absorptions. Besides, the influence of the lifetime 
of the wounded nucleon is not investigated in details before. 
This is the reason why we use the code GCP to study $J/\psi$ 
suppression. We intend to study the influence of the lifetime of 
wounded nucleon on the strength of comover absorption. Our results show
that it is not possible to fit the Pb-Pb data within reasonable choices of  
parameters. Our work is divided into two parts. First, the 
longitudinally excited wounded nucleon
model(LEWNM) is added to GCP. We use it to handle the nucleus-nucleus
interaction. Next, the process of $J/\psi$ production and absorption 
in p-A and A-A collisions in SPS energy is simulated using GCP. 
The results agree well with most experiment 
data of $J/\psi$ production in p-A and A-A collisions, except 
for the data in Pb-Pb collision, where the hadronic 
absorption mechanisms failed to follow the abnormally strong 
$J/\psi$ suppression.

The outline of the paper is as follows: In  section II, we describe 
the transport approach and cascade method.
In section III, the nucleus-nucleus interaction in GCP is described.
In section IV, the nucleon and comover absorption 
mechanisms for  $J/\psi$ suppression is investigated. In the last 
section, some further discussions
are given. \\

\centerline{\large{\bf II. Transport approach and cascade method}}

\medskip
In a many body system, if correlation between particles is weak, i.e. 
independent particle pictures is a good approximation, and if the 
interaction is dominated by short-range forces, we may treat the system
as a collection of point particles propagating freely between 
successive short-range interactions. This system can be described 
by a set of Boltzmann equations, which can be solved using the 
cascade algorithm\cite{high}\cite{gene}.

In the absence of interactions, particles travel on straight line 
trajectories, and the equation for the time evolution of $\omega_a$ is 
\begin{eqnarray}
\label{e1}
p^{\mu}\partial_{\mu}\omega_a(\stackrel{\rightarrow}{x},t,\stackrel
{\rightarrow}{p}) = 0.
\end{eqnarray}
$\omega_a$ is the probability density of finding an on-shell particle 
$a$, with momentum $\stackrel{\rightarrow}{p}$, at position 
$x^\mu=(\stackrel{\rightarrow}{x},t )$. 
Considering the interaction between particles, an $S-$matrix is included
to determine the interaction going from the initial state 
$|b_1, b_2, \cdots, b_n>$ to the final state $|c_1, c_2, \cdots, c_m>$, 
\begin{eqnarray}
\label{e2}
<c_{1},c_{2}, \cdots ,c_{m}|S| b_{1},b_{2}, \cdots ,b_{n}> =
A_{n \rightarrow m}(2 \pi)^4  \delta ^4(\sum \limits_{i=1}^n p_{b_i}
- \sum \limits_{j=1}^m p_{c_j}).
\end{eqnarray}
Summing over all possible incoming channels and outgoing channels,
we get a very general transport equation, with local interactions, 
in the absence of mean fields, 

\begin{eqnarray}
\label{e3}
p^{\mu}\partial_{\mu}\omega_a(\stackrel{\rightarrow}{x},t,\stackrel
{\rightarrow}{p})
&=&  \sum \limits_{n} \sum \limits_{b_1, b_2, \cdots , b_n}
\int\prod\limits_{i=1}^n
 \frac{d^3{\stackrel{ \rightarrow}
 {p}_{b_i}}}{(2 \pi)^3 2E_{b_i}}\omega_{bi}(\stackrel{ \rightarrow}{x},
 t, \stackrel {\rightarrow}{p_{bi}}) \cr
& &\sum \limits_{m} \sum \limits_{c_1, c_2, \cdots , c_m} \int
\prod \limits_{j=1}^m
\frac{d^3{ \stackrel{ \rightarrow}{p}_{c_j}}}{(2\pi)^3 2E_{c_j}}
|A_{n \rightarrow m}|^2 \cr
& &(2 \pi)^4 \delta ^4( \sum \limits_{l=1}^n p_{b_l}- \sum
\limits_{k=1}^mp_{c_k}) \cr
& &[-\sum \limits_{i=1}^n \delta_{ab_i} \delta^3( \stackrel{ \rightarrow}
{p}- \stackrel{ \rightarrow} {p_{b_i}})2E_{b_i} + 
\sum\limits_{j=1}^m \delta_{ac_j} \delta^3(\stackrel{\rightarrow}{p} - 
\stackrel{\rightarrow}{p_{c_j}})2E_{c_j}],
\end{eqnarray}
where '$-$' and '$+$' in the fourth line before $\delta$ functions  
represent the decrease of particles of type 
$a$ from the incoming channels and the increase of $a$ in the
outgoing channels, respectively.
In principle, given the initial values of the distribution and all 
$A_{n \rightarrow m}$'s, Equation(\ref{e3}) can be solved.

A particle cascade is one of the best method for solving these 
equations. In a cascade, the probability distributions $\omega_a$ of
particle $a$ are sampled by $M_a$ test points,
\begin{eqnarray}
\label{e4}
\omega_a(\stackrel{\rightarrow}{x},t,\stackrel{\rightarrow}{p}) = 
\frac{N_a}{M_a}\sum\limits_{i=1}^{M_a} \delta^{3}(\stackrel{\rightarrow}{x}
- \stackrel{\rightarrow}{x}_{i}(t))
\delta^3(\stackrel{\rightarrow}{p} - \stackrel{\rightarrow}{p}_i),
\end{eqnarray}
where $N_a$ is the total number of $a$. 
In the limit $ M_a\rightarrow \infty$, $\omega_a$ can be described 
accurately by the density of test points. In the collision free
limit, 
\begin{eqnarray}
\label{e5}
\omega_a(\stackrel{\rightarrow}{x},t,\stackrel{\rightarrow}{p}) = 
\frac{N_a}{M_a}\sum\limits_{i=1}^{M_a} \delta^{3}(\stackrel{\rightarrow}{x}
- \stackrel{\rightarrow}{x}_{i}(0) - \stackrel{\rightarrow}{v}_i t)
\delta^3(\stackrel{\rightarrow}{p} - \stackrel{\rightarrow}{p}_i)
\end{eqnarray}
is a solution to Eq.(\ref{e1}), where 
$\stackrel{\rightarrow}{v_i} = \stackrel{\rightarrow}{p}_i/ E_i$. 
Between collisions particles travel along a straight line, as in 
Eq.(\ref{e5}). 

For the case of resonance decay (n=1),
\begin{eqnarray}
\label{e6}
\frac{1}{2E_b} \int \prod \limits_{j=1}^m
\frac{d^3{ \stackrel{ \rightarrow}{p}_{c_j}}}{(2\pi)^3 2E_{c_j}}
|A_{1\rightarrow m}|^2 \cr
(2 \pi)^4 \delta ^4(p_{b}- \sum \limits_{k=1}^mp_{c_k}) \cr
= \eta_{b \rightarrow c_1 + c_2 + \cdots + c_m} .
\end{eqnarray}
$\eta_{b \rightarrow c_1 + c_2 + \cdots + c_m}$ represents the decay rate
for the channel of $b \rightarrow c_1 + c_2 + \cdots + c_m$. The total 
decay rate is  $\eta_b = \sum \limits_{m} \sum \limits_{c_1, c_2,
\cdots, c_m}\eta_{b \rightarrow c_1 + c_2 + \cdots + c_m}$.
One can convert the decay rate to the resonance lifetime,  
$\eta_b = 1 / (\gamma_b \tau_b)$, where 
$\gamma_b = |\stackrel{\rightarrow}{p_b}| / E_b$ is the Lorentz decay
factor, $\tau_b$ is the inherent lifetime of resonance state $b$.

In two body collisions (n = 2), 
\begin{eqnarray}
\label{e7}
\frac{1}{2E_{b_1} \cdot 2E_{b_2}}
\int \prod \limits_{j=1}^m
\frac{d^3{ \stackrel{ \rightarrow}{p}_{c_j}}}{(2\pi)^3 2E_{c_j}}
|A_{2\rightarrow m}|^2 \cr
(2 \pi)^4 \delta ^4(p_{b_1} + p_{b_2}- \sum \limits_{k=1}^mp_{c_k}) \cr
= |\stackrel{\rightarrow}{v_2} - \stackrel{\rightarrow}{v_1}|
\sigma_{b_1 + b_2 \rightarrow 
c_1 + c_2 + \cdots + c_m},
\end{eqnarray}
where $\sigma_{b_1 + b_2 \rightarrow c_1 + c_2 + \cdots + c_m}$ 
is the partial cross-section for the channel of 
$b_1 + b_2 \rightarrow c_1 + c_2 + \cdots + c_m$. 
The total cross-section is found by summing over all outgoing
channels, 
\begin{eqnarray}
\label{e8}
\sigma_{b_1 + b_2} = \sum \limits_{m} \sum \limits_{c_1, c_2,
\cdots, c_m}\sigma_{b_1 + b_2 \rightarrow c_1 + c_2 + \cdots + c_m}.
\end{eqnarray}
The product 
$|\stackrel{\rightarrow}{v_2} - \stackrel{\rightarrow}{v_1}|\cdot\sigma_{b_1 + b_2}$ is a cylindrical volume with a length 
$|\stackrel{\rightarrow}{v_2} - \stackrel{\rightarrow}{v_1}|$
 and a cross-section $\sigma_{b_1 + b_2}$. It represents  the 
probability of collision between particles $b_1$ and $b_2$ in a unit time. 
We can simulate this term by making a collision between a point particle 
$b_1$ and a point particle $b_2$ whenever they are approaching each other 
within a cross-section $\sigma_{b_1 + b_2}$. We can set the collision
time to be the time when the distance between the two
particles is at a minimum. When a collision occurs, the
partial cross-sections are used to determine the branching
ratio to a particular channel. 

The momentum distribution for the outgoing particles are selected within
the phase space, weighted by $|A_{n \rightarrow m}|^2$, 
and constrained by the energy momentum conservation\cite{lore},
\begin{eqnarray}
\label{e10}
\int \prod \limits_{j=1}^m
\frac{d^3{ \stackrel{ \rightarrow}{p}_{c_j}}}{(2\pi)^3 2E_{c_j}}
|A_{n \rightarrow m}|^2 \cr
(2 \pi)^4 \delta ^4( \sum \limits_{l=1}^n p_{b_l}- 
\sum \limits_{k=1}^mp_{c_k}). \cr
\end{eqnarray}
The cascade reduces the solution of the transport equations into 
the scattering of a set of classical point particles with known 
cross-sections and branching ratios. 
However, there is one problem unsolved. Although the transport equation is 
local and Lorentz invariant, the cascade breaks this invariance 
by allowing particles to collide at a distance 
$d = \sqrt{\sigma/\pi}$ apart (for n = 2). The collisions are time
ordered and this ordering is frame dependent. Therefore, Lorentz 
invariant is not
strictly satisfied. The codes based on cascade method generally 
have this problem.
The result of Monte Carlo simulation depends on the selection of frame 
more or less. At present energy range, the results of different
frames consistent with each other qualitatively. However, 
the problem will be more serious at RHIC energy region. 
In ref.\cite{high} a method is proposed to recover this invariance. \\

\centerline{\large{\bf III. The nucleus-nucleus interaction in GCP}}

\medskip

In building the interaction of GCP, it is convenient and instructive
to build up the physics step by step.

1. Nucleon-nucleon interaction

In GCP, the processes of the resonance production, $\pi$ production and 
the resonance decay are included in nucleon-nucleon interaction table,
\begin{eqnarray}
\label{e11}
N  +  N  \longrightarrow  N^*   +  N^* ,
\end{eqnarray}
\begin{eqnarray}
\label{e12}
N  +  N  \longrightarrow  N  +  N   +  l \pi,
\end{eqnarray}
\begin{eqnarray}
\label{e13}
N^{*}  \longrightarrow  N  +   \pi, 
\end{eqnarray}
where $l$ is the number of produced $\pi$, which is related to the energy
in the center of mass frame. The probability of the interactions is determined
by the branching ratios. In GCP, the difference between proton(p) and
neutron(n) is not considered. $\pi^+, \pi^- , \pi^0 $ are also regarded
the same as $\pi$.

(1). The process of $\pi$ production 
 
According to the experimental results, the average number of produced 
charged particles in N-N collisions can be expressed as\cite{wong}
\begin{eqnarray}
\label{e14}
<N_{ch}> = 0.88 + 0.44 ln s + 0.118(ln s )^2,
\end{eqnarray}
where $s$ is the square of the energy in the center of mass frame, 
in unit $GeV^2$.
Since the difference of $\pi^+, \pi^-$ and $\pi^0 $ is neglected, after
removing the contribution of leading particles, we can get the 
average multiplicity of pions 
\begin{eqnarray}
\label{e14a}
<n> = [<N_{ch}> - 1.5] * \frac {3}{2}. 
\end{eqnarray}
Introducing the selected 
multiplicity distribution, one can use GCP to get the multiplicity $n$ for
each collision. The energy and momentum conservation are constrained
during the collision processes. No other dynamics mechanism is 
included in GCP. In GCP, we choose $P_t$-limited phase space distribution. 
Considering the leading particle effect of incoming nucleon, 
uniform longitudinal momentum distribution of pions is selected. 
We put KNO distribution into GCP and get the  
multiplicity distribution of pions which is shown in fig.1.
The statistical results of the rapidity distribution and the transverse 
momentum distribution of the produced particles in N-N collisions is shown 
in fig.2.

(2). The process of resonance decay

$N^*$ is a resonance state which is produced during N-N interaction. 
In GCP, $N^*$ will decay to N and $\pi$ within a given time $\tau$, 
which is shown in Eq.(\ref{e13}).

2. p-A and A-A collision

The description that $\pi$ is produced immediately after each N-N 
collision in p-A and A-A collisions is not actually the case. 
The longitudinally excited wounded nucleon model(LEWNM) is 
introduced in GCP. In LEWNM, each N-N collision leads to two 
longitudinally excited clusters. The clusters decay after a certain time. 
The decay of clusters is the source of other produced particles in 
N-N collision. 

In LEWNM the nucleons are not transversely excited but only 
longitudinally stretched after the N-N collision. The collision of a beam 
nucleon 
and a target nucleon  will result in two excited clusters, $'string1'$ 
and $'string2'$:
\begin{eqnarray}
\label{e20}
& &      N  +  N  \longrightarrow  string_1  +  string_2.
\end{eqnarray} 
The mass of the clusters are fixed in the following way: By using GCP, 
the process of $\pi$ production in N-N collisions is simulated. In the 
center of mass frame, the particles are divided into two clusters according
to their $z$ direction. The invariant masses of all particles in 
each of the two clusters are counted over a broad range of c.m. energies 
of the N-N system. 
Then we can get the empirical formula of the mass of the cluster 
which is a function of the c.m. energy of the N-N system.

The longitudinally excited clusters $'string1'$ and $'string2'$ subsequently
decay,  
\begin{eqnarray}
\label{e20a}
string   \longrightarrow  N  +  l \pi. 
\end{eqnarray}
If the resonance or string collide with other particles  before 
decaying  new excited clusters may produce.
Decay of the clusters are the source of particle production. 

The secondary collisions are also considered in p-A and A-A collision.
The processes mainly include:
\begin{eqnarray}
\label{e21}
N(N^*, string)  +  \pi  \longrightarrow  N  +   \pi  + l \pi, 
\end{eqnarray}
\begin{eqnarray}
\label{e22}
\pi  +  \pi  \longrightarrow  \pi  +   \pi  +  l \pi .
\end{eqnarray}
The process of N-N collision is not changed after introducing 
the LEWNM into GCP. According to the experimental condition, 
we also have simulated the S-U collision. The
statistical results of transverse energy($E_T$) distribution 
are shown in figure 3. 
The LEWNM gives a good description of many observable 
quantities in N-N, p-A and A-A collisions. It can be 
applied to the study of relativistic nucleus-nucleus reactions.

3. Collision geometry 

The nuclear geometry plays an important role in relativistic heavy ion
collisions. The overlapping area of the two colliding nuclei is 
determined by the impact parameter $b$.
We call the nucleons in the overlapping area the participants. 
The number of participants  is 
calculated according to the nuclear geometry.
Considering the change of $b$, we get the distribution of the 
number of participants.
The larger the $b$, the less the number of participants. In GCP, $b$ is an 
input parameter. It can be fixed or selected randomly  in a certain region.

In section II, we have discussed that the collision is possible when the 
distance of closest approach of the two colliding particles 
is less than the interaction range $d$, 
related to the total cross-section by $d = \sqrt{\sigma/\pi}$.
The possible future collisions are ordered in time and form the collision 
list. The next collision is the earliest one on the list.
 
So the basic considerations of GCP only relate to collision 
geometry and cascade simulation, no other dynamics of the production 
process is included.  The aspects we mentioned 
above are only related to the nuclear geometry and kinematics. 
The detailed physical assumptions can be introduced according to the 
physical models. This is the remarkable advantage of GCP.\\ 
\\

\centerline{\large{\bf IV. The hardronic mechanisms on $J/\psi$ 
suppression}}

\medskip 
Two hardronic mechanisms of $J/\psi$ suppression are added to
GCP to simulate the production and absorption of $J/\psi$.

1. The process of $J/\psi$ production and absorption  

The channels for $J/\psi$ production and absorption are added  to
GCP to simulate the process of $J/\psi$ production and absorption:  

\begin{eqnarray}
\label{e15}
N(N^*, string) + N(N^*, string) \longrightarrow N + N + J/\psi
\end{eqnarray}
\begin{eqnarray}
\label{e17}
N(N^*, string) +  J/\psi   \longrightarrow  N  +  D  +  \bar D
\end{eqnarray}
\begin{eqnarray}
\label{e18}
\pi  +  J/\psi   \longrightarrow  D  +   \bar D
\end{eqnarray}
Since the probability of $J/\psi$ production in N-N collision is 
relatively small, the cross-section of $J/\psi$ production is enlarged 
in the simulation to increase $J/\psi$ production probability. In general, 
one or two $J/\psi$ are produced in each event in the simulation.


2. The mechanism of the $J/\psi$ absorption in nuclear matter

Based on above discussions, one can accept the following physical
picture that the $J/\psi$ production can be divided into two steps.
The first step is the production of a $c \bar c$ pair, which is produced 
perturbatively and almost instantaneously. The second step is the 
formation of a physical state of $J/\psi$, that needs a much longer
time. 
In a nucleus-nucleus collision, $c \bar c$ are produced by hard 
scattering processes of beam nucleon and one target nucleon. The 
produced $c \bar c$ may interact with another target nucleon and these
$c \bar c$-nucleon interactions may lead to the break up of $c \bar c$
via the reaction
\begin{eqnarray}
\label{e23}
c\bar c  +  N(N^*, string) \longrightarrow D + \bar D + x ,
\end{eqnarray}  
which turns $c \bar c$ into $D \bar D$ pair. Thus, $c \bar c$-nucleon
interactions will give  a suppression of $J/\psi$ production.

3. The mechanism of the interaction between $J/\psi$ and produced mesons 

Comovers usually refer to the secondaries produced in high energy 
heavy ion collisions, such as $\pi,  \rho$ and $\omega$ mesons, etc. 
In A-A collisions, besides $J/\psi$-nucleon absorption, 
$J/\psi$ particles also suffer interaction with secondaries that 
happen to travel along with them, which also causes $J/\psi$ 
suppression. The reaction is shown in Eq.(\ref{e18}) .

The details of simulating the process of $J/\psi$ production 
and absorption are as following:

(1). An entry for $J/\psi$ production is added to the interaction table.
We use average impact parameter $<b>$ to discuss minimum biased data. 
For the situation of different $E_T$ bin the values of b are 
given by experimental groups\cite{NA38}\cite{NA50}. In order to increase the 
probability of $J/\psi$ 
production, we enlarge the cross-section of $J/\psi$ production to 
ensure one or two $J/\psi$ produced in every event. Running GCP,
we get statistical number of $J/\psi$, $N_{pro}$, in 10000 events.

(2). Put the channel for $J/\psi$-nucleon interaction into GCP. Using
the same condition as (1), we run GCP again to count 
 the number of $J/\psi$, $N_{abs(N)}$, in 10000 events, where 
 the nucleon absorption is included. 

(3). At last, $J/\psi$-comover interaction is added to GCP. Using the
same condition as (1), we obtain the number of $J/\psi$, $N_{abs(N + co)}$,  
after considering the nucleon and comover absorption.  

The $J/\psi$ survival probability in A-B collision is expressed as 
\begin{eqnarray}
\label{e24}
S = N_{abs} / N_{pro} .
\end{eqnarray}

The absorption cross sections of $J/\psi$, i.e. the cross-sections 
that the produced $J/\psi$ particles interact with target nucleons or 
produced secondary particles, are
important parameters to explain the $J/\psi$ suppression in hardronic
environment. There are two parameters, $J/\psi$-nucleon cross-section
$\sigma_{abs(N)}$ and $J/\psi$-comover cross-section $\sigma_{abs(co)}$.
After analyzing many sets of experimental data in p-A collisions, 
Gerchel and H$\ddot{u}$fner\cite{gerc} found that the experimental 
$J/\psi$ production 
data for p-A collisions can be fitted well with an effective 
$J/\psi$-nucleon cross section $\sigma = 6.2 mb$ or $6.9mb$.
To account for the A-A data, $J/\psi$-comover absorption
cross-section is generally regarded as about $3mb$.
In GCP, $\sigma_{abs(N)}$ and $\sigma_{abs(co)}$ are input parameters.
We wish to adjust the parameters to agree well with the experimental data
of $J/\psi$ suppression. 

First we discuss the case of minimum biased data. 
Considering only the absorption of
$J/\psi$ by nucleus, the GCP simulation of $J/\psi$ survival 
probabilities at SPS energy are expressed as squares in fig.4. The 
experimental data are shown as black triangles. The $J/\psi$-nucleon 
cross-section is taken to be $\sigma_{abs(N)} = 7mb$, which is in 
good agreement to earlier works\cite{gerc}. The lifetime of the wounded 
nucleon is taken to be 1fm/c.  
One can see clearly that our simulation is in good agreement 
with the experiment data of $J/\psi$ production in p-A collisions. 
Most of the $J/\psi$ suppression data for A-A collisions are also fitted, but 
one can not explain the data in Pb-Pb collision. 
Using GCP, we can simulate the $J/\psi$ 
production and absorption in p-A and A-A collisions up to S-U data 
successfully. Next, $J/\psi$-comover interaction is added to GCP. 
Considering the 
nucleon and comover absorptions of $J/\psi$, we repeat the above procedure  
again. The value of $\sigma_{abs(N)}$ is taken the same as above. 
The $J/\psi$-comover absorption cross-section is $\sigma_{abs(co)} = 2mb$.  
The results are expressed as open triangles in fig.4. It shows that the  
$J/\psi$-comover interaction in A-A collisions is more important 
than that in p-A collisions. However, to include the comover 
absorption has not introduced qualitative difference from the result 
including only nucleon absorption. The combination of these two absorption 
mechanisms of $J/\psi$ still can not explain the data for Pb-Pb 
collision. In the above situation, the lifetime of the wounded nucleon 
is taken as 1fm/c. In GCP, the decay of the wounded nucleons is the source 
of particle production. If their 
lifetime decreases, other particles such as mesons will be 
produced earlier. This may increase the probability 
of $J/\psi$-comover interaction. Fig.5 shows the results 
in p-A and A-A collisions obtained using the lifetime of the wounded
nucleons  as 
0.2fm/c. It can be seen from the result that the shorter lifetime 
does provide a much stronger comover absorption. 
Now the Pb-Pb data can be reached. However the other fitting points are 
much lower than the experimental data in lighter A-A collisions. 
The above analysis tell us that one can not explain the 
$J/\psi$ suppression for all the observed experimental data consistently 
based on the one set of parameters. 
Then we adjust $\sigma_{abs(N)}$ to 6mb, 
$\sigma_{abs(co)}$ to 3mb, repeat the above step again. In this 
case, the comover absorption of $J/\psi$ is more manifest. But 
the overall results are similar to fig.4.         

Now we turn to discuss the result of different $E_T$ bins. In the same
way as above, we simulate the two kinds of 
$J/\psi$ suppression,  the interaction of $J/\psi$ with nuclear 
matter and produced meson in S-U and Pb-Pb collisions. 
For different $E_T$ bins, the values of b are given by 
experimental groups, which are listed in table 1. 
The parameters $\sigma_{abs(N)}$ and $\sigma_{abs(co)}$ used in 
considering the $J/\psi$-nucleon absorption and $J/\psi$-comover 
absorption and the lifetime of the wounded nucleon are taken to 
be the same as those used for fig.4. in fitting 
the minimum biased data. The squares in fig.6 
show that $J/\psi$ suppression in S-U collision could be described 
very well based on our nucleon absorption. However, the fitting 
for the last $E_T$-bin in Pb-Pb collision using only nucleon 
absorption shows that (the most right square in fig.6) the data in Pb-Pb 
collision could not be fitted using the same set of parameters. 
The interaction of $J/\psi$ with produced mesons make 
further suppression of $J/\psi$ in S-U and Pb-Pb collisions, which 
are shown as triangles. 
But the combination of these two kinds of absorption 
(see ``$\bigtriangledown$ '' in fig.6) 
still can not explain the extra strong $J/\psi$ 
suppression in Pb-Pb collision. There may be anomalous suppression 
caused by other mechanisms, e.g. QGP formation. \\

\centerline{\large{\bf V. Results and Discussions}}

\medskip
GCP is designed such that the cascade algorithm is separated from the 
physics models. In this way the problem of the cascade can be 
isolated and studied separately, and different physics models can be 
compared in a common framework. GCP itself is only a tool to simulate
the cascade process. The various physics models can be built based on
it. This is the 
key difference between GCP and most other cascade codes.

In this paper, we focus our attention on the Monte Carlo simulation 
of $J/\psi$ production and absorption directly by using GCP.
Our work is divided into two parts. First, the longitudinally 
excited wounded nucleon model(LEWNM) is added to GCP. 
It is used to handle the nucleus-nucleus interaction. The LEWNM is 
found to give a good description of many observables  in N-N and p-A
collisions, such as the multiplicity distributions, the rapidity 
distributions, and the transverse momentum distributions 
of the produced particles. According to the experimental conditions, 
we have also simulated the transverse energy distribution 
for S-U collision.  
Next, two hardronic mechanisms of $J/\psi$ suppression,  
$J/\psi$-nucleon interaction and $J/\psi$-comover interaction, are 
added into GCP to simulate the production and absorption of $J/\psi$
in p-A and A-A collisions at SPS energy. First we consider the 
case of minimal biased data. Then the results of different 
$E_T$ bins are discussed. 
Using the nucleon absorption mechanism, the results agree well 
with the experimental data of $J/\psi$ production, except 
for those in Pb-Pb collision. Based on above discussion, 
the comover absorption
mechanism is added to GCP to study the suppression of $J/\psi$ 
production. The $J/\psi$ suppression in Pb-Pb collision can not
be explained even by the combination of the above two mechanisms.
It seems that some new mechanisms is  needed to study the 
anomalous $J/\psi$ suppression in Pb-Pb collision.
This may indicate the formation of QGP.  

\vskip 5mm

\centerline{\Large{\bf Acknowledgments}}

\vskip 5mm
This work is supported in part by the National
Natural Science Foundation of China and Natural Science
Foundation of HeBei province. We would like to thank 
Zhang X.F. and Tai A. for stimulating and helpful discussions.

\newpage

\vskip 1cm
\centerline{\Large{\bf Figure Caption}}
\vskip 5mm
Fig.1: the statistical results of the multiplicity distribution 
of pions in N-N collision.
\vskip 5mm

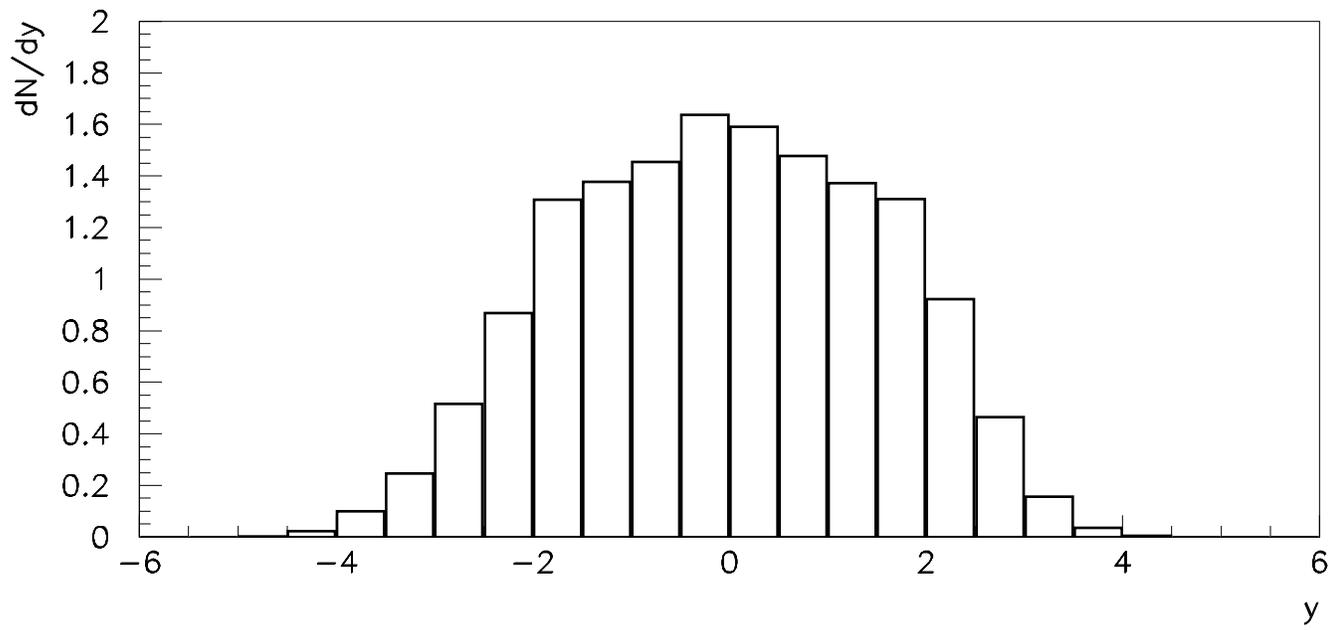
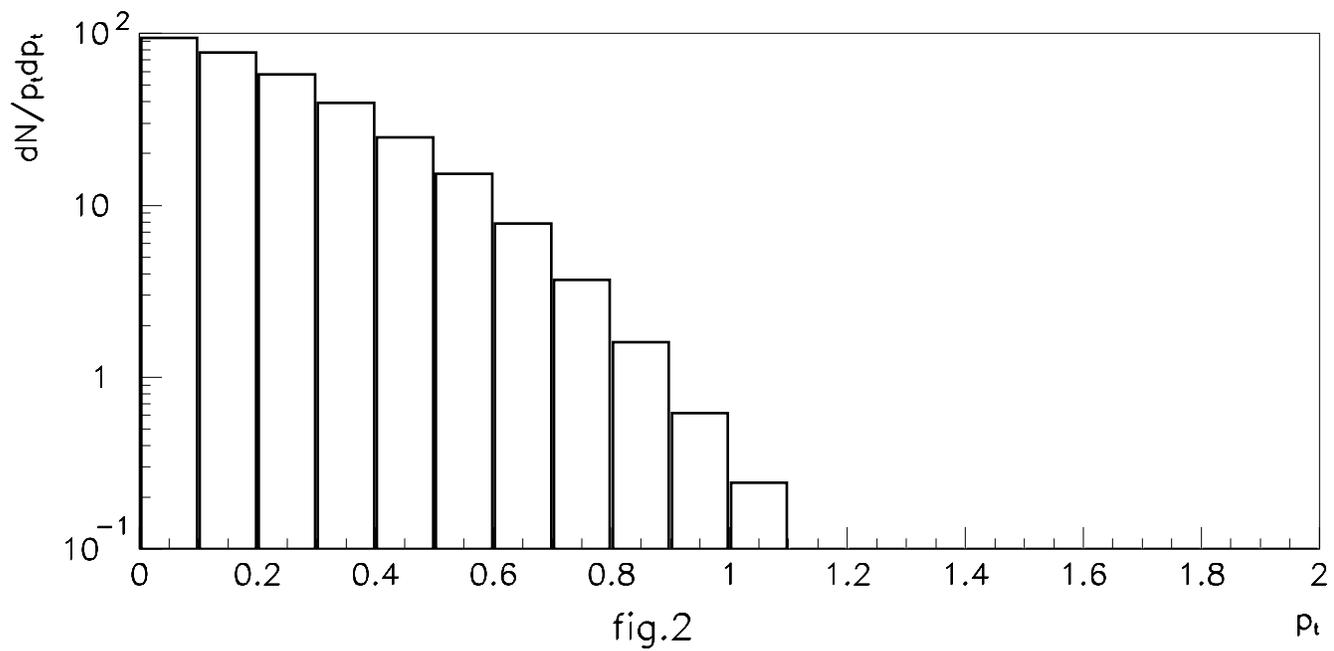
Fig.2: the statistical results of the rapidity distribution and the 
transverse momentum distribution of the produced particles 
in N-N collision.
\vskip 5mm

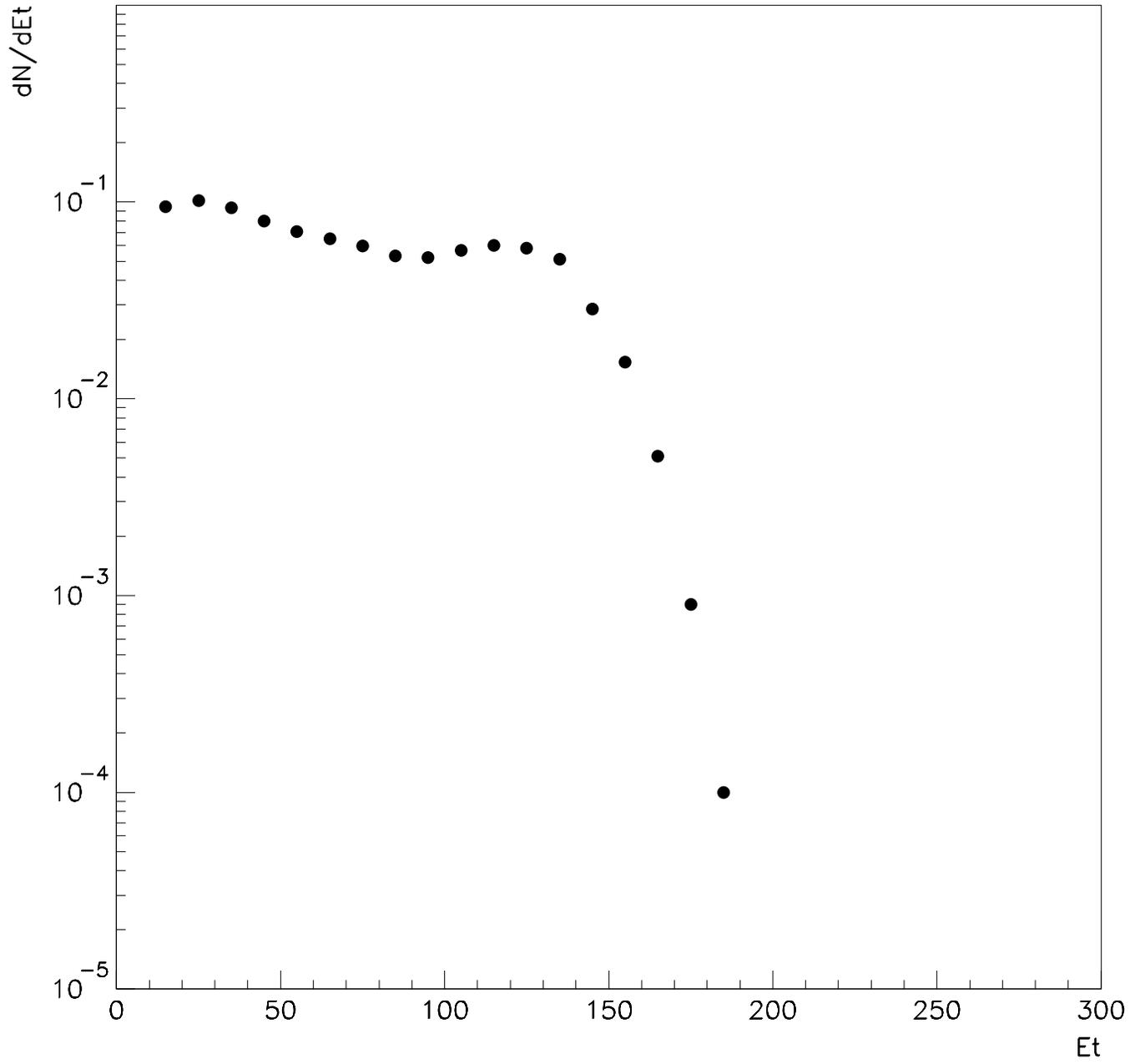
Fig.3: the statistical results of the transverse energy 
distribution of the produced particles 
in S-U collision.
\vskip 5mm

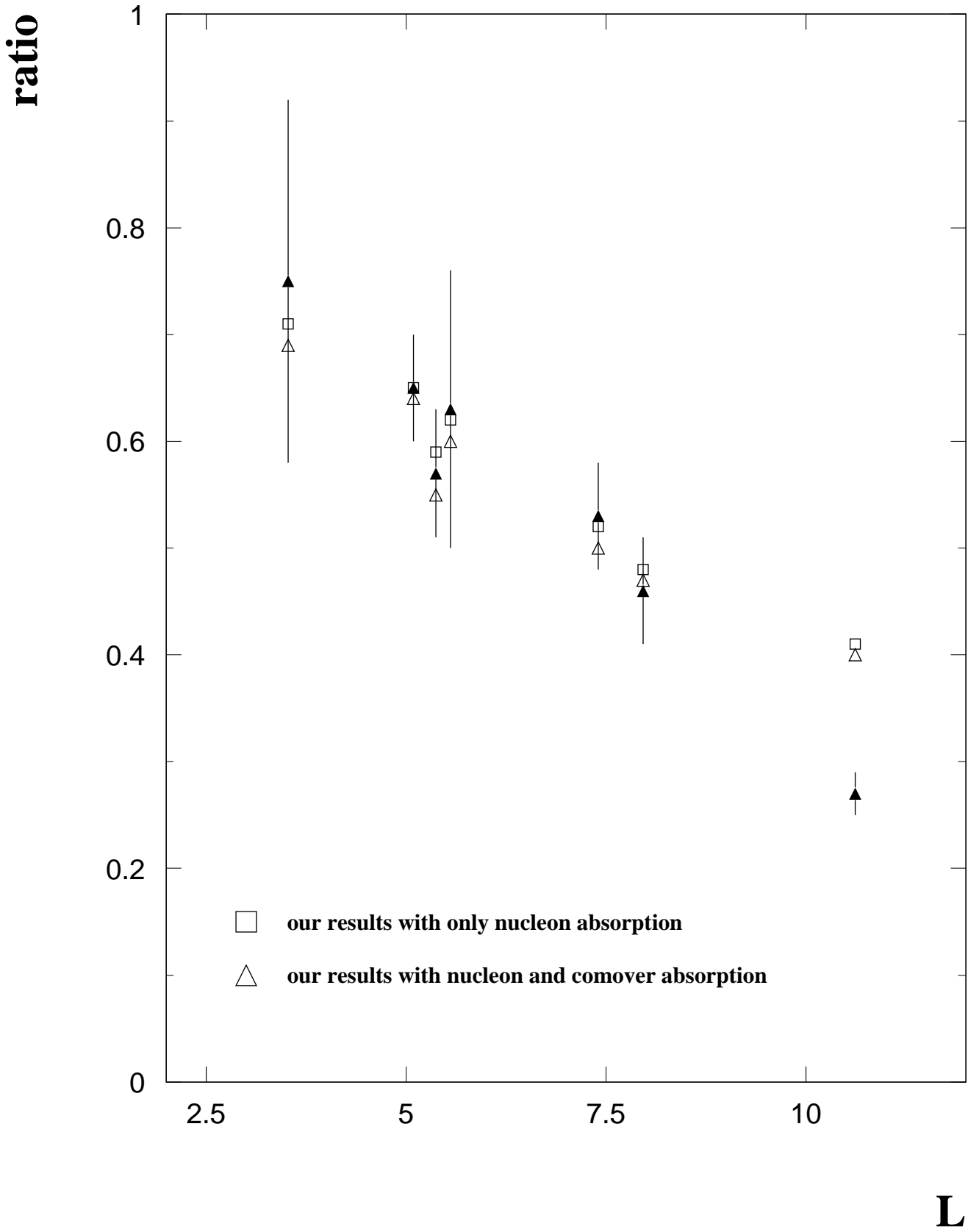
Fig.4: The $J/\psi$ survival probability in p-A and A-A collisions 
obtained using the lifetime of the wounded nucleon as 1fm/c, together 
with minimum biased experimental data for(from left to right) 
p-Cu, p-W, p-U, O-Cu, O-U, S-U and Pb-Pb collisions . 
\vskip 5mm

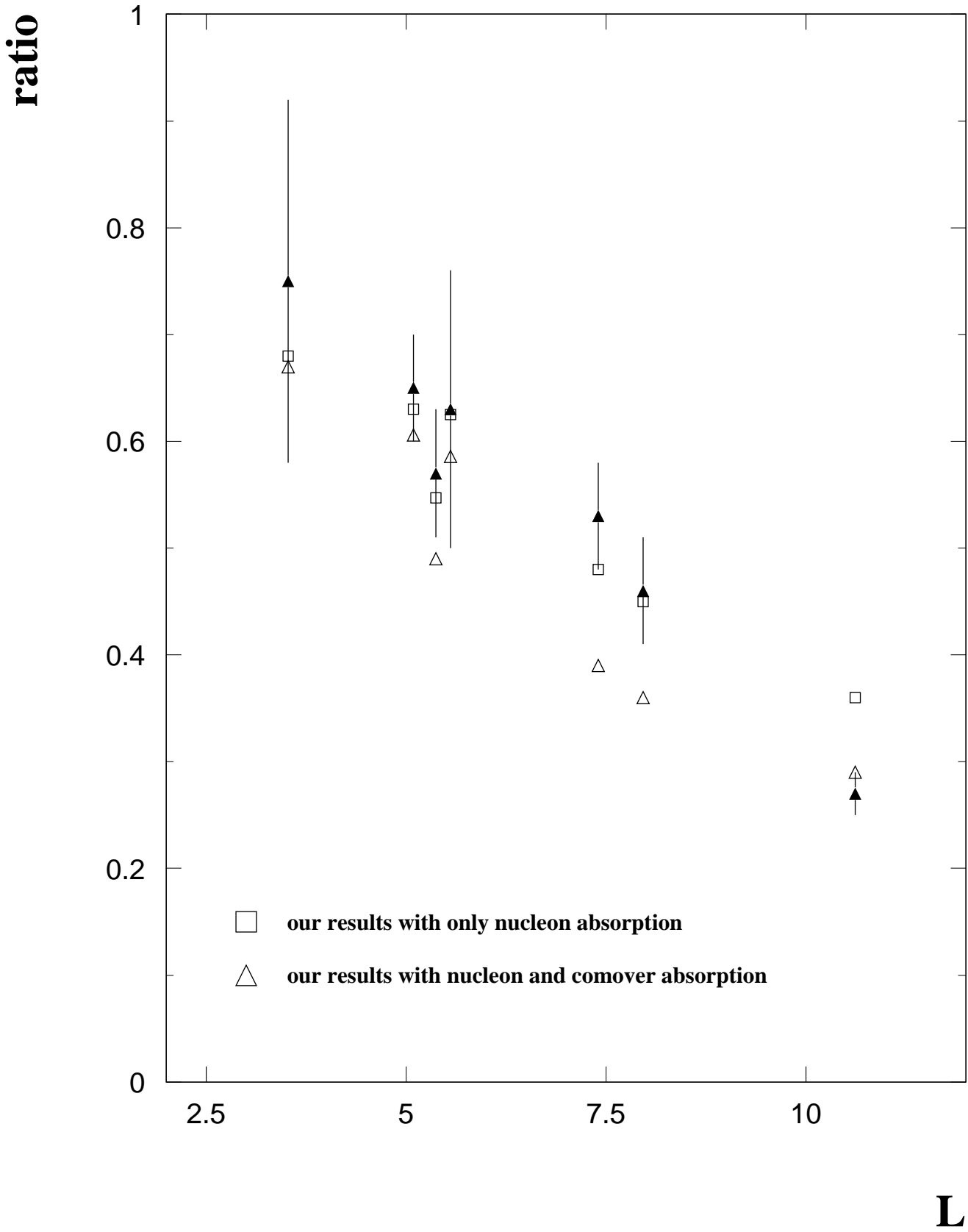
Fig.5: The same as Fig.4 using the lifetime of wounded nucleon 
as 0.2fm/c. 
\vskip 5mm

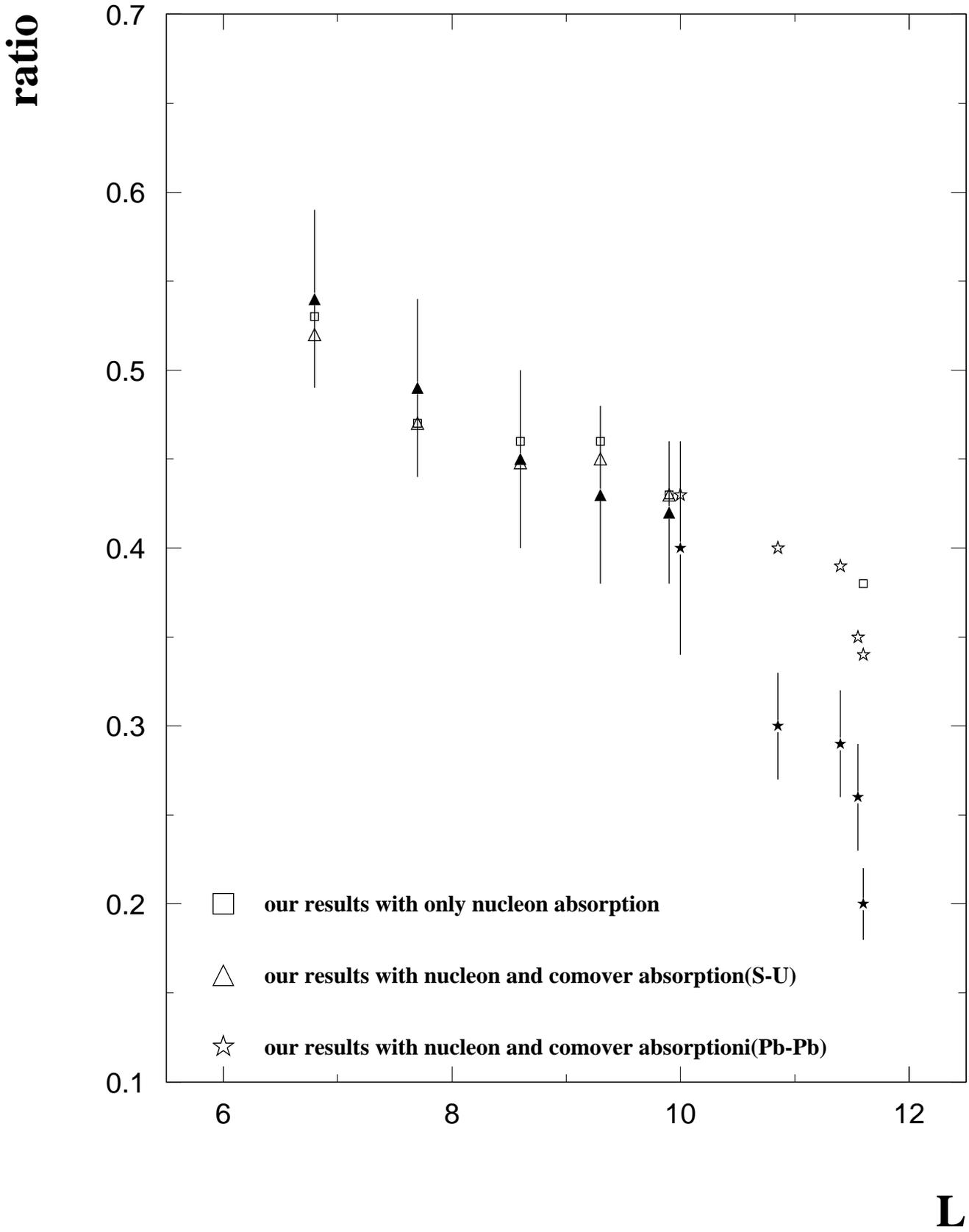
Fig.6: The $J/\psi$ survival probability for different $E_T$ bins
together with experimental data in S-U and Pb-Pb collisions.
\vskip 5mm

\vskip 1cm
\centerline{\Large{\bf Table Caption}}
\vskip 5mm

Table I. The value of $<b>$ for different $E_T$ bins in A-A collisions.

\newpage
\begin{center}

Table I

\vskip 1cm

\tabcolsep 0.3in
\begin{tabular}{|c|c|c|c|c|c|c|c|c|} \hline \hline
 A-A  &            & bin1  & bin2  & bin3  &  bin4 & bin5  \\ \cline{1-7}
 S-U  & $E_T$      & 34    & 58    & 88    &  120  & 147   \\ \cline{2-7} 
      & $<b(E_T)>$ & 7.2   & 5.5   & 4.4   &  3.6  & 2.4   \\ \cline{1-7}
Pb-Pb & $E_T$      & 25    & 42    & 57    &  71   & 82    \\ \cline{2-7}  
      & $<b(E_T)>$ & 8.3   & 6.8   & 5.0   &  3.3  & 1.8   \\ \hline
\end{tabular}
\end{center}
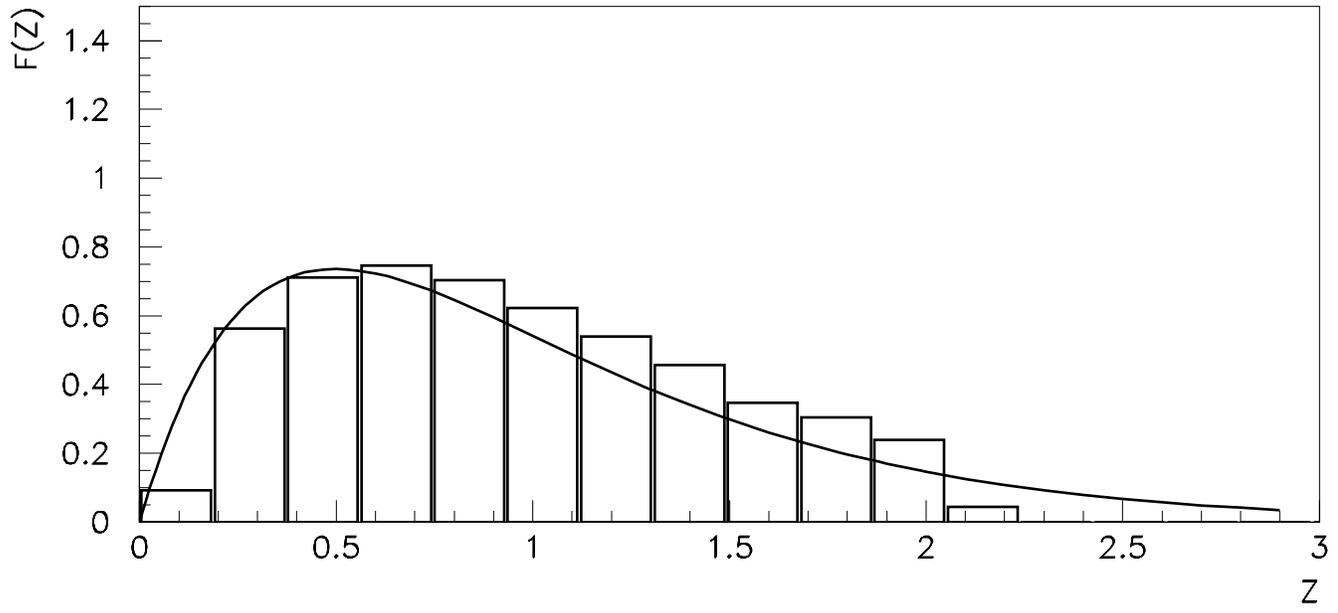
\end{document}